\documentclass[12pt]{article}
\usepackage{amssymb}
\usepackage{graphicx}
\usepackage[cp1251]{inputenc}
\usepackage{rotating}
\begin{document}
 \bigskip
 \centerline{\bf FEATURES OF THE DISTRIBUTION OF ABSORBING MATTER}
 \centerline{\bf IN THE LOCAL SYSTEM}
\bigskip
 \bigskip
 \centerline{\bf
            V. I. Sorokina$^1$\footnote [99]{e-mail: vasilisushka05@gmail.com},
            V. V. Bobylev$^2$,
            G. A. Gontcharov$^{2}$,
            A. T. Bajkova$^2$
            }
 \bigskip
   \centerline {\small \it $^1$Saint Petersburg State University, Saint Petersburg, Russia}
   \centerline {\small \it $^2$Main (Pulkovo) Astronomical Observatory of the Russian Academy of Sciences, St. Petersburg, Russia}
 \bigskip
 \bigskip
A detailed study of the dust distribution in the Local System was conducted. Using the latest map by Gontcharov et al., smoothed distributions of dust matter were obtained in projection onto the galactic plane $XY$ using various smoothing parameters. Within the 2-kpc radius region around the Sun under study, key structural features associated with the Radcliffe Wave, Split, Sagittarius Spur Extension, Malpolon+Natrix, and Vela Ridge superclouds are clearly identified. It was shown that the Radcliffe Wave, Malpolon+Natrix, and Vela Ridge regions exhibit periodic perturbations of vertical coordinates with wavelengths ranging from 2.5 kpc (Radcliffe Wave) to 2 kpc (Malpolon+Natrix and Vela Ridge). No similar long-wavelength, high-amplitude oscillations of vertical coordinates were detected in the Sagittarius Spur Extension and Split regions.


\newpage
\section{INTRODUCTION}
Between the two segments of the spiral arms closest to the Sun, namely, Perseus and Carina-Sagittarius, is located the Local System (Local Arm, Orion Arm). This structure has been known for a long time (Mignet, 1930; Ogorodnikov, 1965; Efremov,
 1989). It consists of gas and dust clouds, HII zones, young massive OB stars, low-mass T Tauri stars, OB associations, young open star clusters, etc. The spatial, kinematic and dynamic properties of the Local System are studied, for example, by Olano (2001), Bobylev, Baikova (2014), Vallee (2018), Veselova, Nikiforov (2020), Xu et al. (2013; 2023).

It is generally accepted that the Orion Arm is a spur of the Grand Design spiral arm.
However, it is currently difficult to determine which arm it is branching off from—Perseus or Carina-Sagittarius—as more precise distances to the stars are needed to provide a definitive answer. According to estimates by various authors (see the table in Bobylev et al. 2025), the twist angle of the Orion Arm is approximately 10$^\circ$, which is in good agreement with twist angles found for other spiral arms in the Galaxy.

Recently, a number of interesting new structural features have been identified in the Local System. In particular, in an analysis of a three-dimensional dust distribution map in the Local Arm and adjacent regions, Lallement et al. (2019) described a spur (split) over 2 kpc long, extending between the Local Arm and the Carina-Sagittarius Arm, oriented relative to the $Y$ axis at an angle of approximately 45$^\circ$. Furthermore, in an analysis of a large sample of molecular clouds with high-precision distances, Alves et al. (2020) identified the Radcliffe Wave for the first time. This is a thin chain of clouds approximately 2.7 kpc long, located between the Sun and the Perseus Arm at an inclination to the $Y$ axis of approximately 30$^\circ$. The main feature of the Radcliffe wave is the wave-like behavior of the vertical coordinates with a maximum amplitude of deviation from the plane of symmetry of the Galaxy of about 150 pc.

\begin{figure}[t]
{ \begin{center}
  \includegraphics[width=0.6\textwidth]{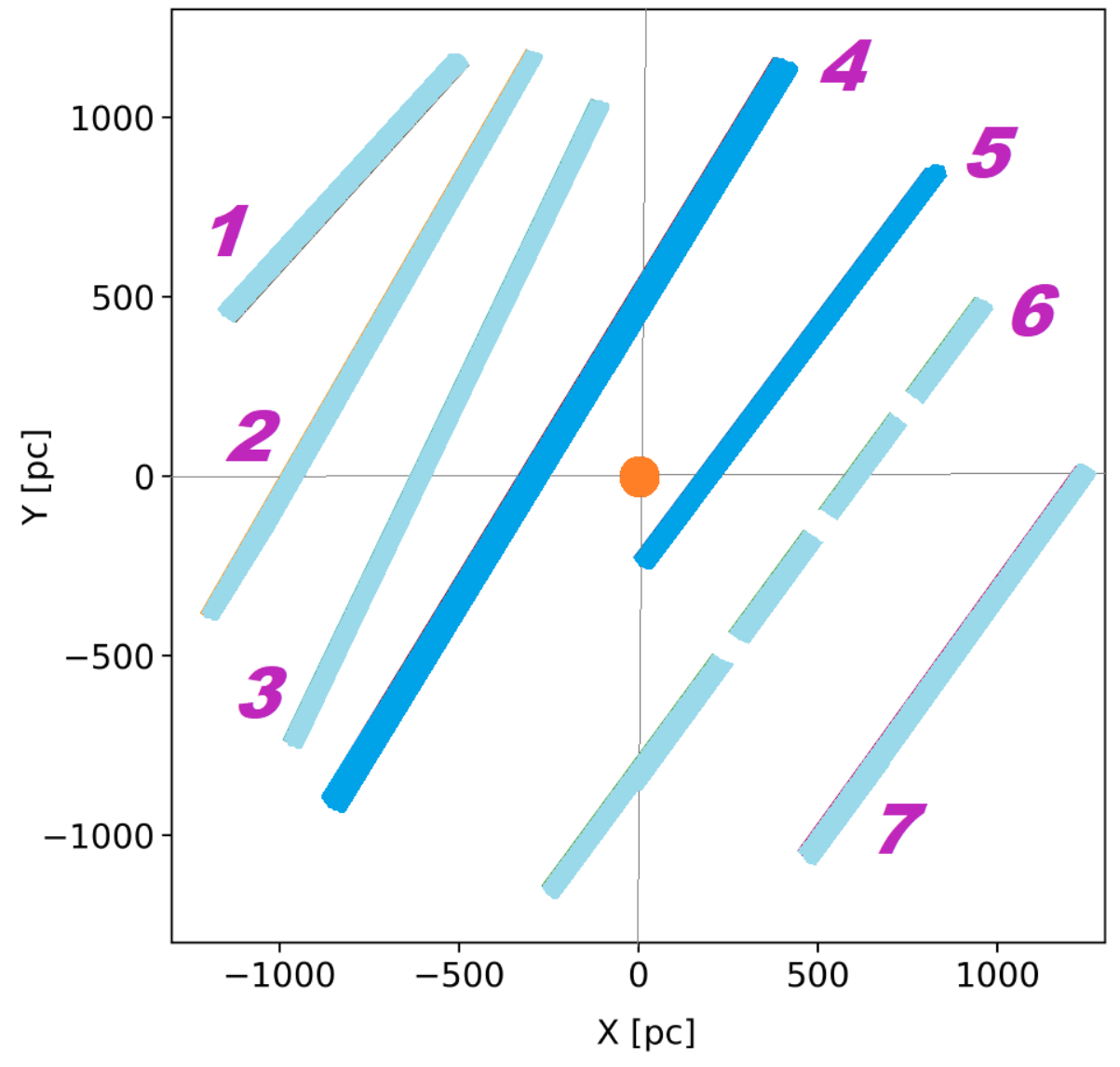}
  \caption{The diagram of dust superclouds in the Local System, adapted by us according to Fig.~4 from the work of Kormann et al. (2026), the following objects are indicated by numbers: 1--Anguis, 2--Malpolon, 3--Natrix, 4--Radcliffe Wave, 5--Split, 6--Vela Ridge and 7--Sagittarius Spur Extension, the Sun, indicated by an orange circle, is located at the origin of the coordinate system.}
 \label{f-1}
\end{center}}
\end{figure}

Finally, in the work of Cormann et al. (2026), as a result of processing the three-dimensional dust distribution map of Edenhofer et al. (2024), an entire grouping of dust superclouds was identified in the Local System. It included such formations as Anguis, Malpolon, Natrix, Radcliffe Wave, Split, Vela Ridge and Sagittarius Spur Extension (see Fig.~\ref{f-1}). Note that most of these superclouds have a ragged structure with gaps. Therefore, in Fig.~\ref{f-1}, Radcliffe Wave and Split are shown in a darker tone as the least ragged (in the map of Edenhofer et al., these two structures are clearly visible without any smoothing). Vela Ridge is indicated by a line with gaps, since it has the most ragged structure. It is interesting to note that in four superclouds, namely Malpolon, Natrix, Radcliffe Wave and Vela Ridge, Kormann et al. (2026) found periodic oscillations of vertical coordinates with wavelengths from 1 to 3 kpc and amplitudes from 30 to 90 pc.

It can be seen that dust distribution maps currently play a key role in studying the fine structure of the Local System. The methods for constructing such maps have both similarities and differences. The main common feature of modern maps is the use of high-precision data from the Gaia project (Gaia Collaboration 2016).

In Lallement et al. (2019), photometric data from the Gaia~DR2 catalog (Gaia 2018 Collaboration) were combined with 2MASS~ measurements (Skrutskie et al. 2006) to obtain extinction near stars with accurate photometry and relative trigonometric parallax errors less than 20\%. This resulted in a three-dimensional dust distribution over a volume of $6\times6\times0.8$~kpc$^3$ around the Sun.
In Edenhofer et al. (2024), distance and extinction estimates for 54 million nearby stars, obtained from Gaia BP/RP spectra, were used to construct a high-resolution 3D map of the interstellar dust distribution.

It is interesting to note the work of Barbillon et al. (2025), who used a highly uniform method based on the spectral chemical-physical parameterization of stars from the Gaia General Stellar Parametriser from Spectroscopy (GSP--Spec) catalog to construct a three-dimensional map of the interstellar dust distribution. According to Barbillon et al., this catalog, consisting of 5.6 million sources in Gaia~DR3 (Gaia 2023 Collaboration), has the advantage of estimating stellar atmospheric parameters independently of extinction. The absorption E(BP--RP) here was calculated from a comparison of the observed color indices (BP--RP) in the Gaia bands with the theoretical ones found from the dependence $T_{\rm eff}-\log(g)$--[M/H]--color, obtained on the basis of the parameters of the stellar atmosphere GSP--Spec ($T_{\rm eff}$ --- effective temperature, $g$ --- acceleration of gravity on the surface of the star, [M/H] --- metallicity, the abundance of $\alpha$-elements relative to iron was also taken into account).

New maps (two- and three-dimensional) of dust distribution are presented in the paper by Goncharov et al. (2025). These maps are based on estimates of $A_V$ and photo-astrometric distances obtained from the dataset of Anders et al. (2022), which uses trigonometric parallaxes from the Gaia~DR3 catalog and multi-band photometry for nearly 100 million dwarf stars. Custom corrections were applied to account for systematic errors in this dataset.

The aim of this paper is to study the structure of the Local System using data from the latest map by Gontcharov et al. (2025). Of greatest interest is the confirmation of periodic long-wave oscillations of vertical coordinates in the regions of the Radcliffe Wave, Malpolon, Natrix, and Vela Ridge superclouds,

\section{DATA}
In the paper by Gontcharov et al. (2025), a two-dimensional (2D) map of interstellar extinction through the entire dust layer in the Galaxy was constructed for galactic latitudes $|b|>13^\circ$ and a three-dimensional (3D) map of extinction within a radius of 2 kpc from the Sun.
The 2D map has an angular resolution of 6.1 arcminutes, while the 3D map has a transverse resolution of 3.56 pc, corresponding to a variable angular resolution depending on the distance, and a radial resolution of 50 pc.

\begin{figure}[t]
{ \begin{center}
  \includegraphics[width=0.99\textwidth]{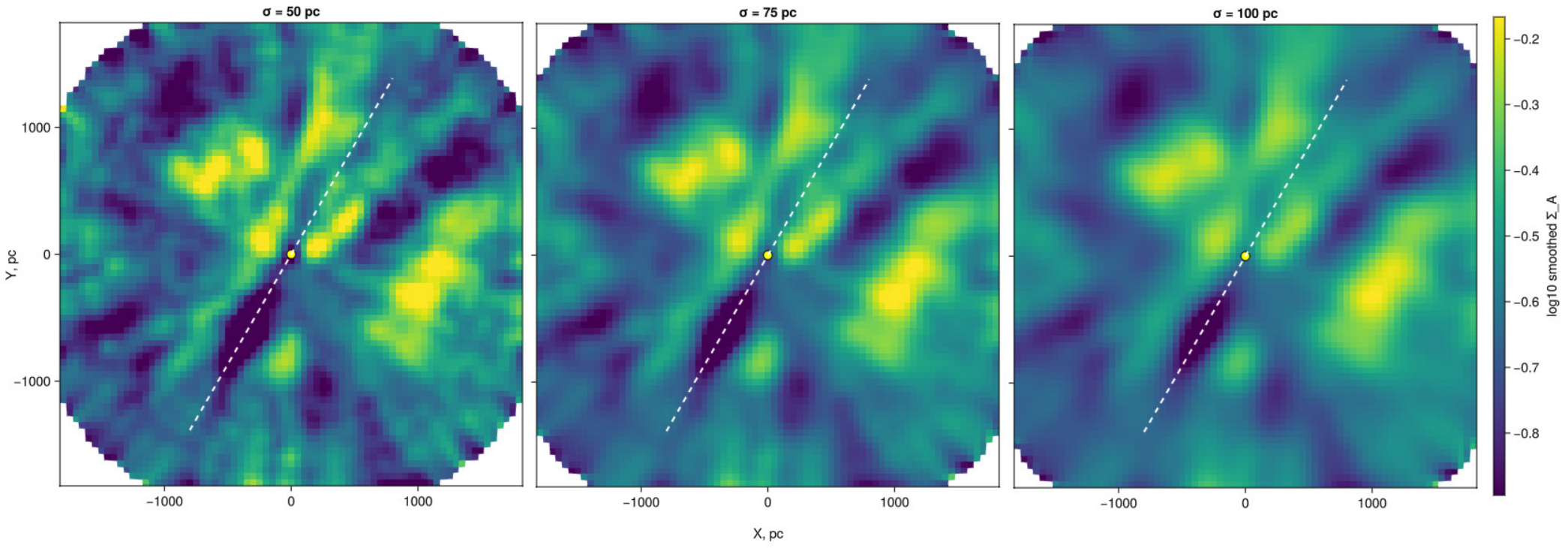}
  \caption{Smoothed distributions of absorbing matter in the region of the Local System, obtained in this paper using the map of Gontcharov et al. (2025) for three cases of $\sigma$, where the stars were selected under the condition $|Z|<300$~pc, the Sun is located at the origin of the coordinate system, the dashed white line is drawn at an angle of 35$^\circ$ to the $Y$ axis.
  }
 \label{f-xy-map}
\end{center}}
\end{figure}

In constructing these maps, special attention was paid to the solar neighborhood (within 200 pc) and high galactic latitudes. The two-dimensional map provides estimates of the total extinction $A_V$ in the galactic dust layer or for extended extragalactic objects beyond this layer with an accuracy of about $\sigma(A_V)=0.07$, and the three-dimensional map provides estimates of $A_V$ for extended objects located within the galactic dust layer with an accuracy of about $\sigma(A_V)=0.1$. For point objects, the estimates may be significantly less accurate due to spatial fluctuations in the dust medium that are not taken into account by the maps. The large dust clouds considered in this study are undoubtedly extended objects. Accordingly, the map predictions for them are quite accurate. The systematic accuracy of these maps can be indirectly estimated from the fact that, during their creation, all systematic effects in the original data, manifested at a level above $\Delta A_V\approx0.03$, were apparently detected, analyzed, and taken into account. If such declared accuracy of the maps is confirmed by further studies, they should be recognized as the most accurate modern extinction maps, respectively, for the three-dimensional map -- over the entire sky within a radius of 2 kpc, and for the two-dimensional map -- at high and middle latitudes ($|b|>13^\circ$) up to distances of many megaparsecs. This conclusion can be made taking into account the limitations and systematic errors of the previously created three-dimensional maps by Lallement et al. (2019), Green et al. (2019) and others, discovered, for example, by Gontcharov, Mosenkov (2017, 2018, 2019, 2021a, 2021b).

Note that Gontcharov et al. (2025) presented extinction estimates based on the maps used here for various classes of extended objects with angular sizes predominantly in the range 2'$–$40', including 19,809 galaxies and quasars, 170 galactic globular clusters, 458 open clusters, and several hundred molecular clouds from two lists, as well as for 8,293 Type Ia supernovae.

\begin{figure}[t]
{ \begin{center}
  \includegraphics[width=0.6\textwidth]{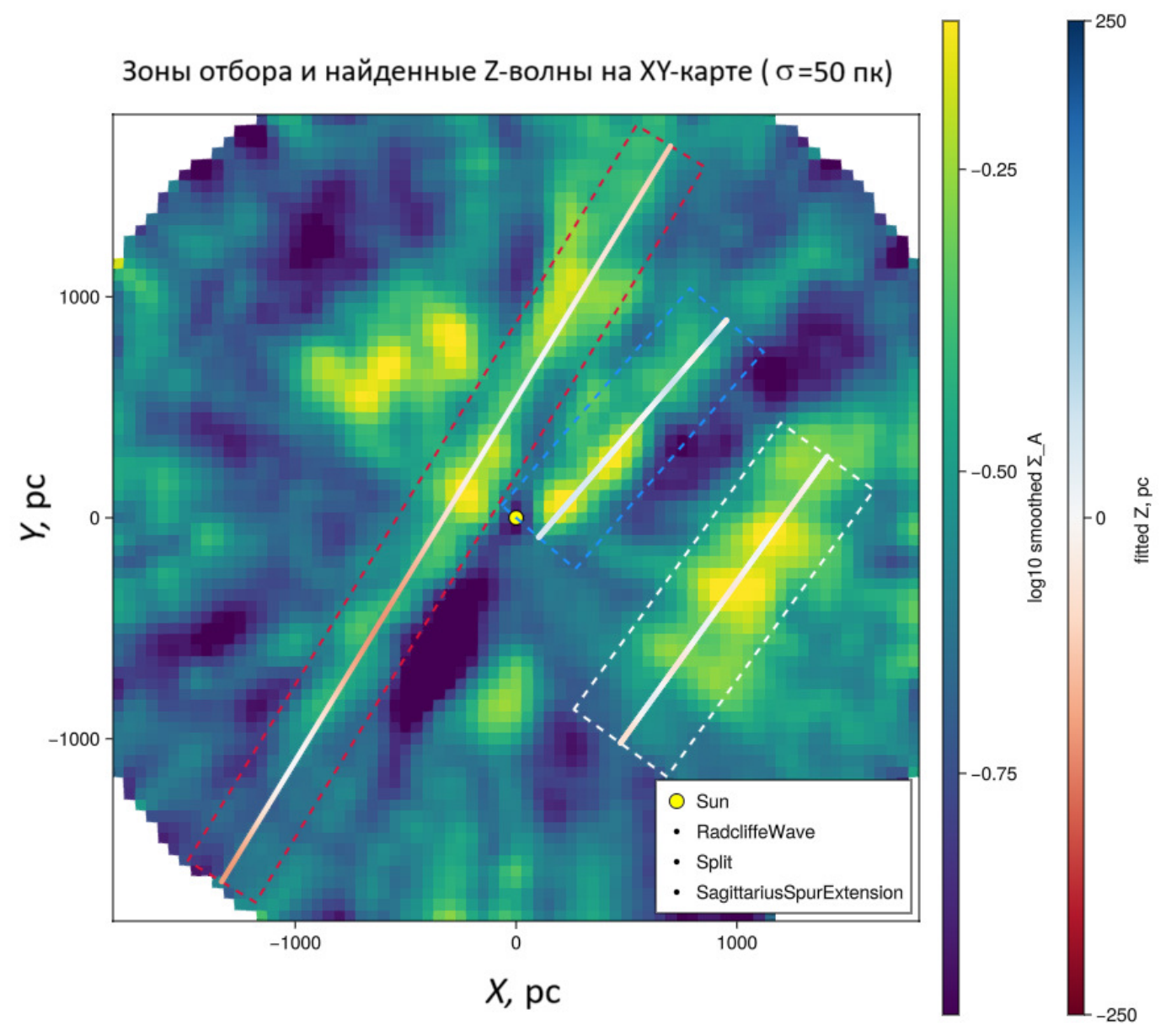}
  \caption{Three selection zones, one color scale on the right shows the depth of the map, and the second scale shows the depth of vertical waves.  }
 \label{f-xy-3}
\end{center}}
\end{figure}
\begin{figure}[t]
{ \begin{center}
  \includegraphics[width=0.85\textwidth]{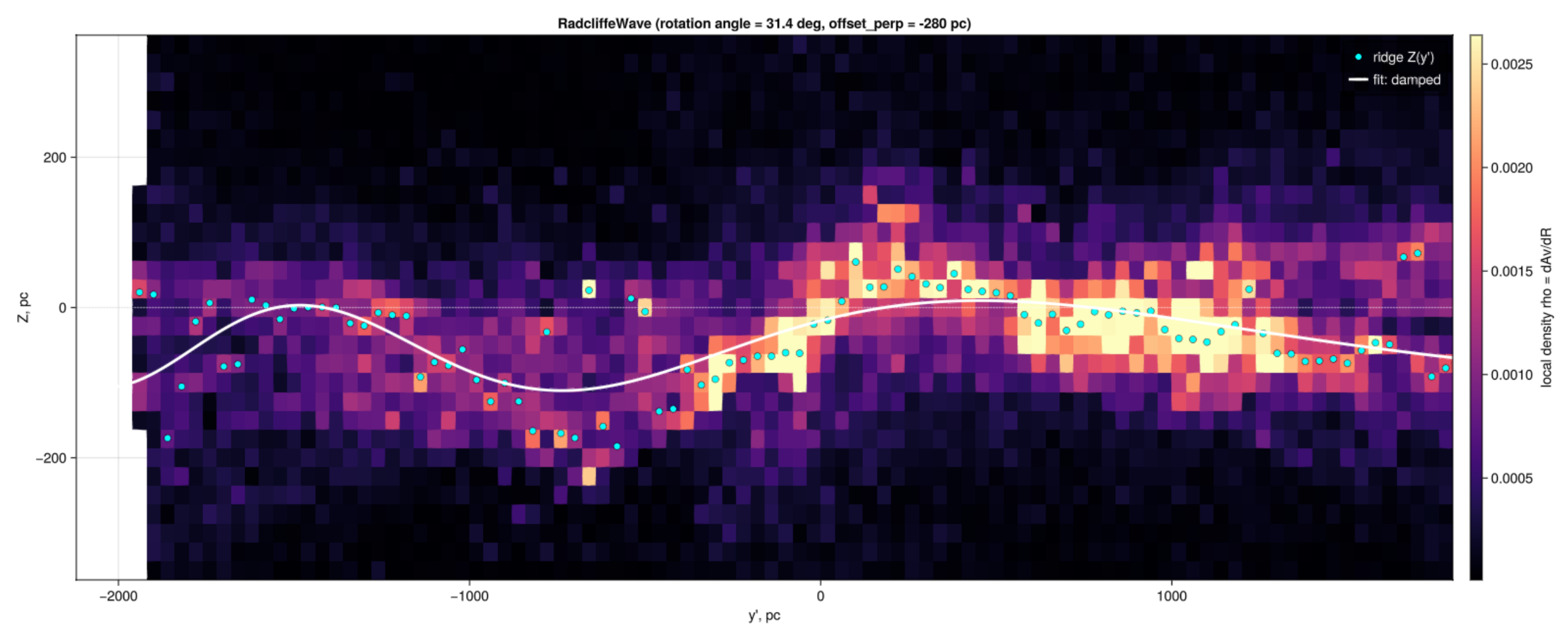}
  \caption{Map of the vertical distribution of dust in the Radcliffe Wave sampling zone. }
 \label{f-xy-4-RW}
\end{center}}
\end{figure}
\begin{figure}[t]
{ \begin{center}
  \includegraphics[width=0.6\textwidth]{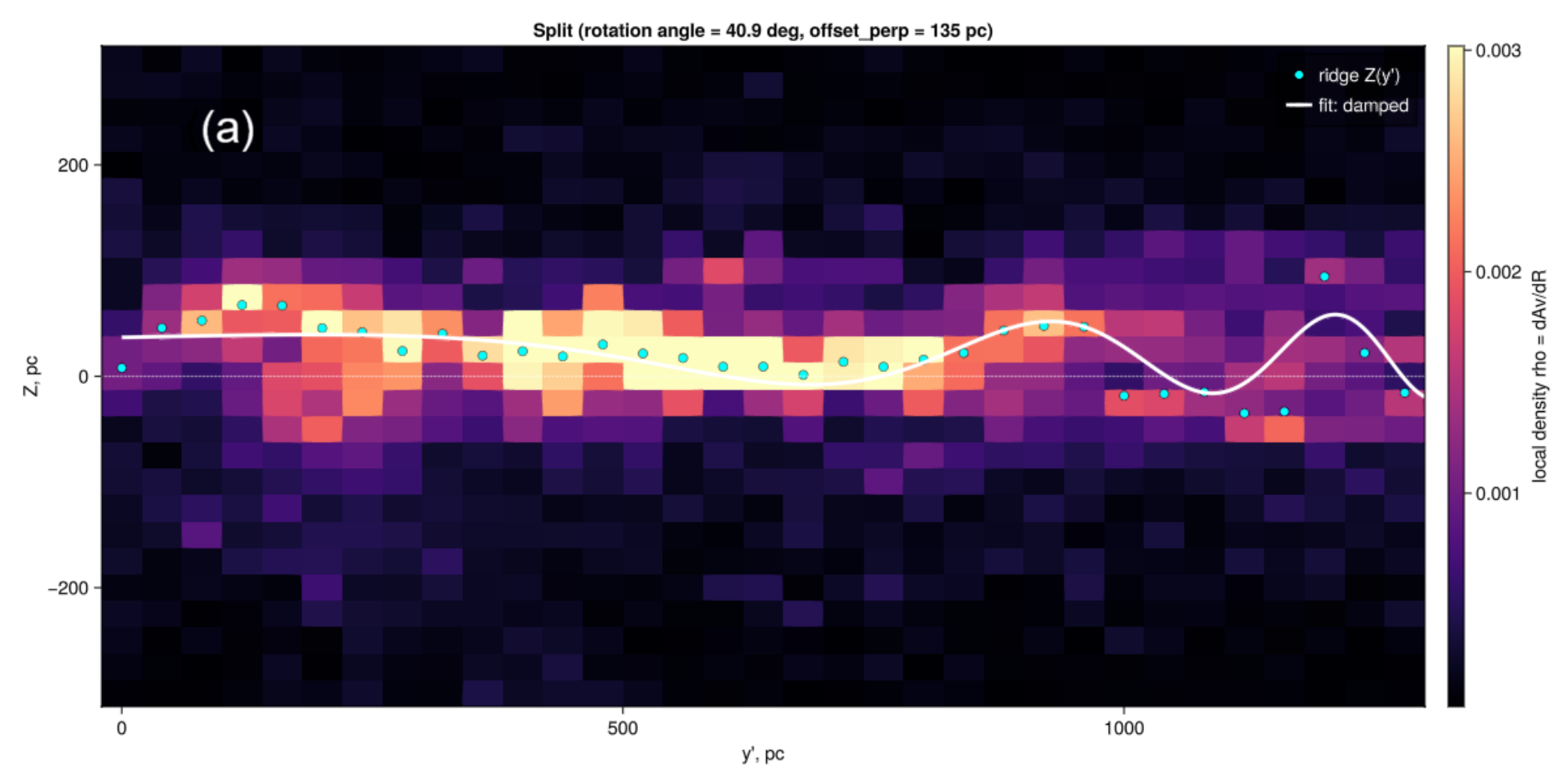}

   \includegraphics[width=0.6\textwidth]{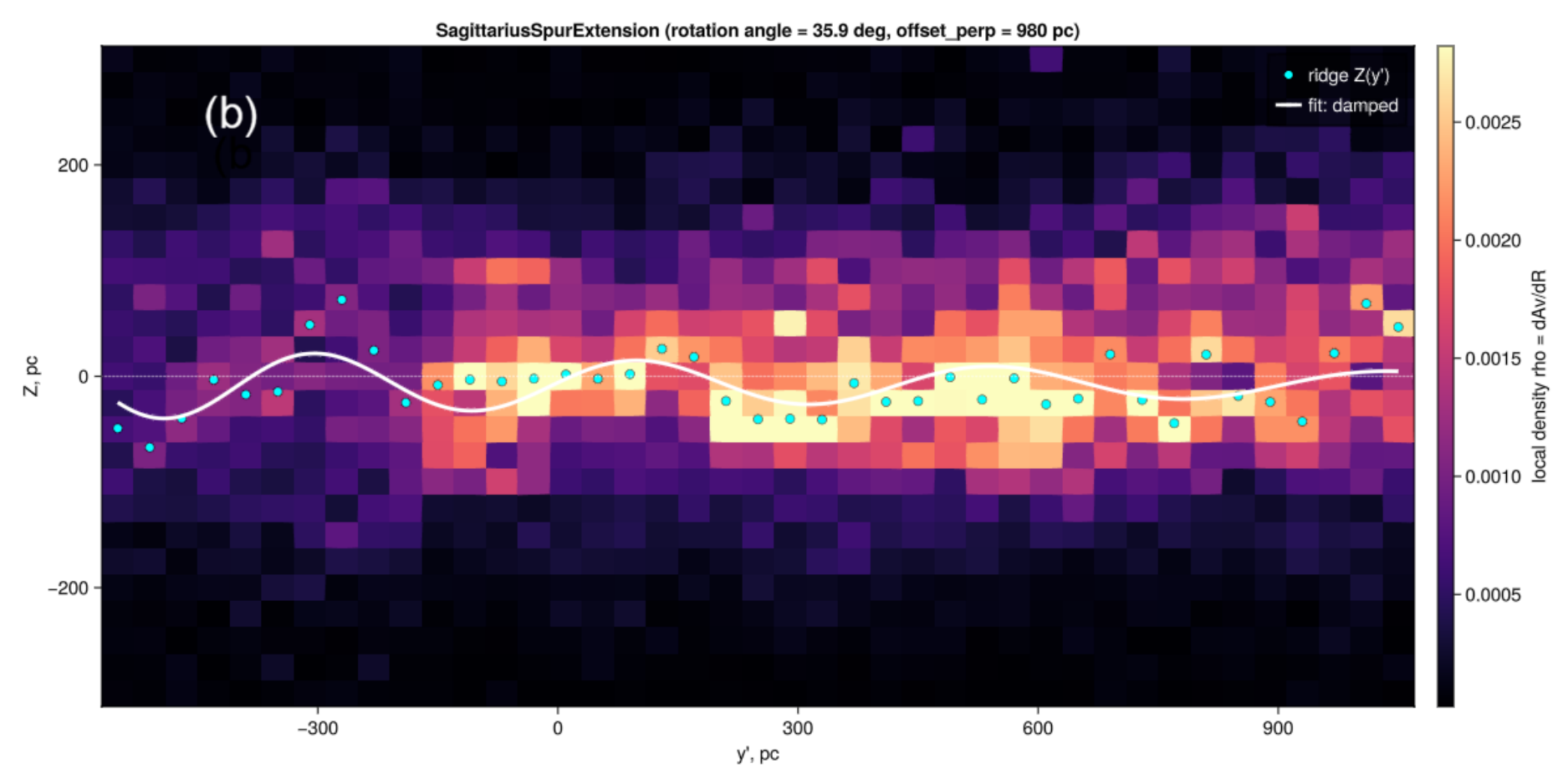}
\caption{Maps of the vertical distribution of dust in the selection zones of the Split (a) and Sagittarius Spur Extension (b) superclouds. }
 \label{f-xy-5}
\end{center}}
\end{figure}

It is also interesting to note the recent work by Zhang et al. (2026), which presents new dust distribution maps constructed using red cluster stars with heliocentric distances down to 7 kpc in the galactic plane ($|Z|$ < 25~pc). The maps have a resolution of 10, 50, and 100 pc. They reveal detailed structures, including spiral arms, interarm branches, and giant cavities. The exponential scale length of the galactic dust disk found from these maps is 2.90 kpc. A region of 2 kpc radius around the Sun contains features useful for solving our problems.

 \section{METHOD}\label{method}
At each point of the heliocentric rectangular grid $(X,Y,Z)$, where the $X$ axis is directed toward the center of the Galaxy, $Y$ --- in the direction of rotation, and $Z$ --- toward the north galactic pole, the corresponding galactic coordinates were calculated

  \begin{equation}
 \begin{array}{lll}
  R=(X^2+Y^2+Z^2)^{1/2},\\
     l=\arctan(Y/X),\\
     b=\arcsin(Z/R).
 \label{eq-1}
 \end{array}
 \end{equation}
The value of $A_V(l,b,R)$ was found by interpolation from the original three-dimensional map. Since the map specifies the integrated extinction, the local value proportional to the dust density was determined by the finite difference

  \begin{equation}
  \begin{array}{lll}
 \rho_A\simeq  [A_V(l,b,R+\Delta R/2)- \\  \qquad -A_V(l,b,R-\Delta R/2)] / \Delta R,\\
 \Delta R=100~\hbox{pc},
 \label{eq-2}
 \end{array}
 \end{equation}
where points outside the radial range of the map were excluded, and negative values were assumed to be equal to zero.

  To construct the projection onto the $XY$ plane, the value $\rho_A(X,Y,Z)$ was summed over the layer $|Z|<300$~pc with a step $\Delta Z=50$~pc. The resulting map $\Sigma_A(X,Y)$ was smoothed by a Gaussian kernel with $\sigma=50$, 75, and 100~pc. The mask of valid nodes $M(X,Y)$ was taken into account during smoothing:
   \begin{equation}
  \begin{array}{lll}
 \widetilde{\Sigma}_A= G_\sigma [M\Sigma_A]   / (G_\sigma M).
  \label{eq-3}
  \end{array}
 \end{equation}
Fig.~\ref{f-xy-map} shows the value of $\log_{10}\widetilde{\Sigma}_A(X,Y)$.

To analyze the disturbances of vertical coordinates in the structures under consideration, a rotated system was used
  \begin{equation}
  \begin{array}{lll}
y'=Y\cos\alpha+X\sin\alpha,\\
 x_\perp=-Y\sin\alpha+X\cos\alpha,
  \label{eq-4}
 \end{array}
 \end{equation}
 where $y'$ is directed along the selected strip.

For each column $y'$, the position of the ridge $Z_r(y')$ was determined from the maximum of the profile $\overline{\rho}_A(Z)$ and refined as a weighted centroid in a window of $\pm60$~pc around the maximum; uninformative columns were discarded. Then, the dependence $Z_r(y')$ was approximated by the function of Kormann et al. (2026)
  \begin{equation}
   \begin{array}{lll}
Z_{\rm fit}=a\exp(\varepsilon_a y')\times \\
  \quad  \times \sin\{\omega\exp(\varepsilon_\omega y')y'+\varphi\}+d.
 \label{eq-5}
  \end{array}
 \end{equation}

 If the attenuation parameters became unstable, a simple sine function $Z_{\rm fit}=a\sin(\omega y'+\varphi)+d$ was used. The fit was performed using weighted least squares, where the weight was the summed density in column $y'$.

\begin{figure}[t]
{ \begin{center}
 \includegraphics[width=0.6\textwidth]{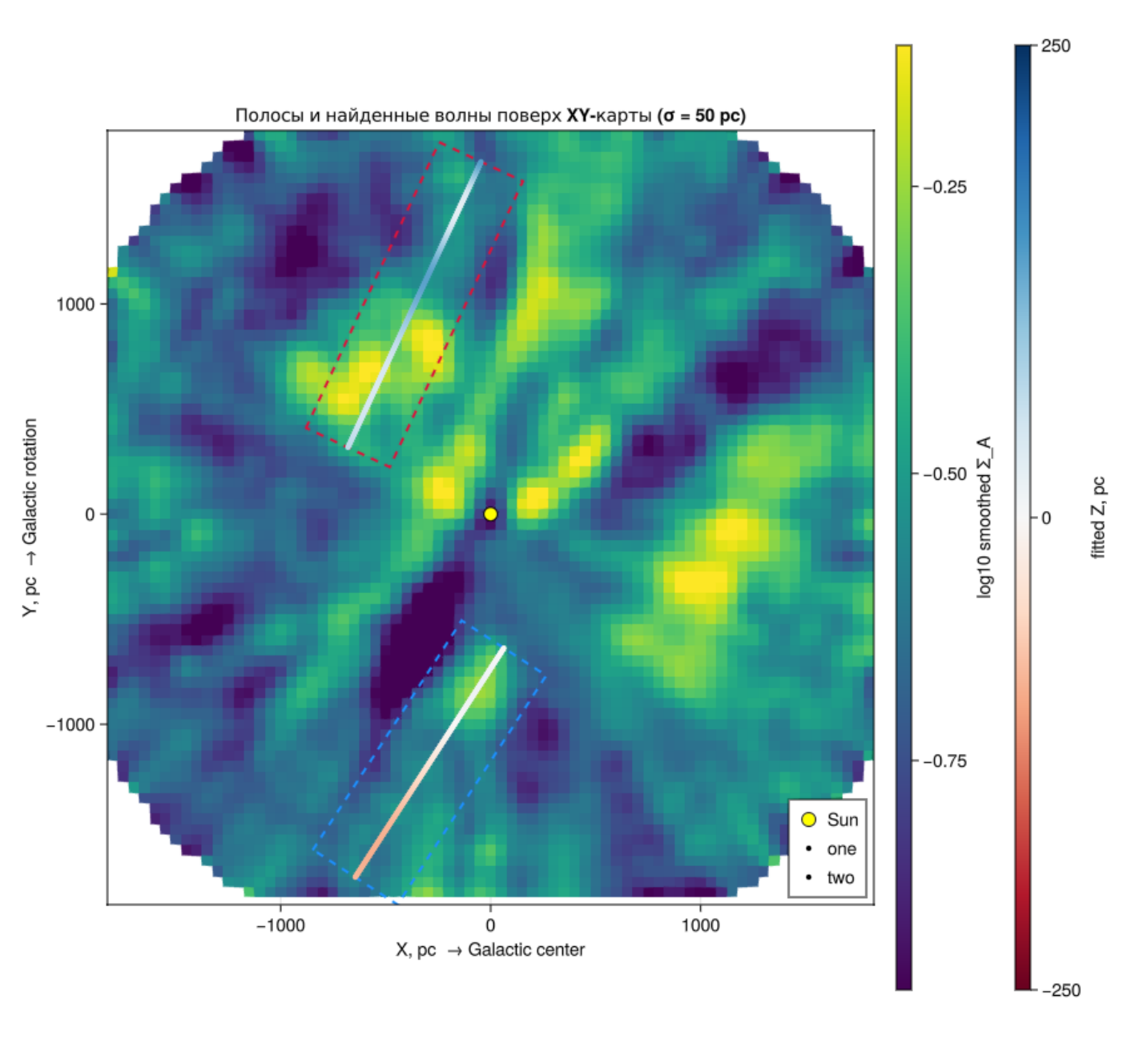}
  \caption{Selection zones in the Malpolon+Natrix area (top left) and Vela Ridge (bottom right), one color scale on the right shows the map depth, and the second scale shows the vertical wave depth.
  }
 \label{f-xy-66}
\end{center}}
\end{figure}
\begin{figure}[t]
{ \begin{center}
  \includegraphics[width=0.6\textwidth]{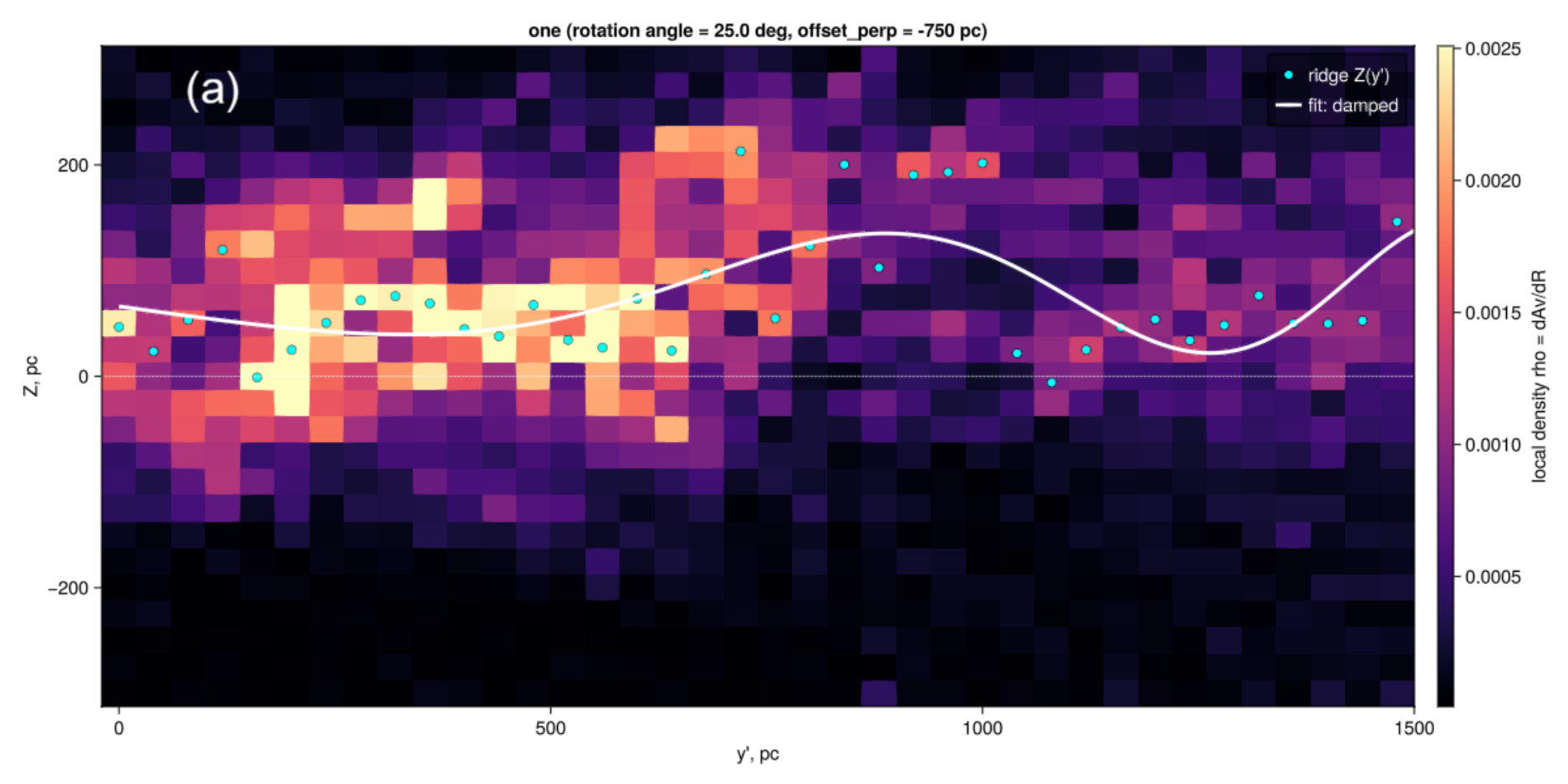}

  \includegraphics[width=0.6\textwidth]{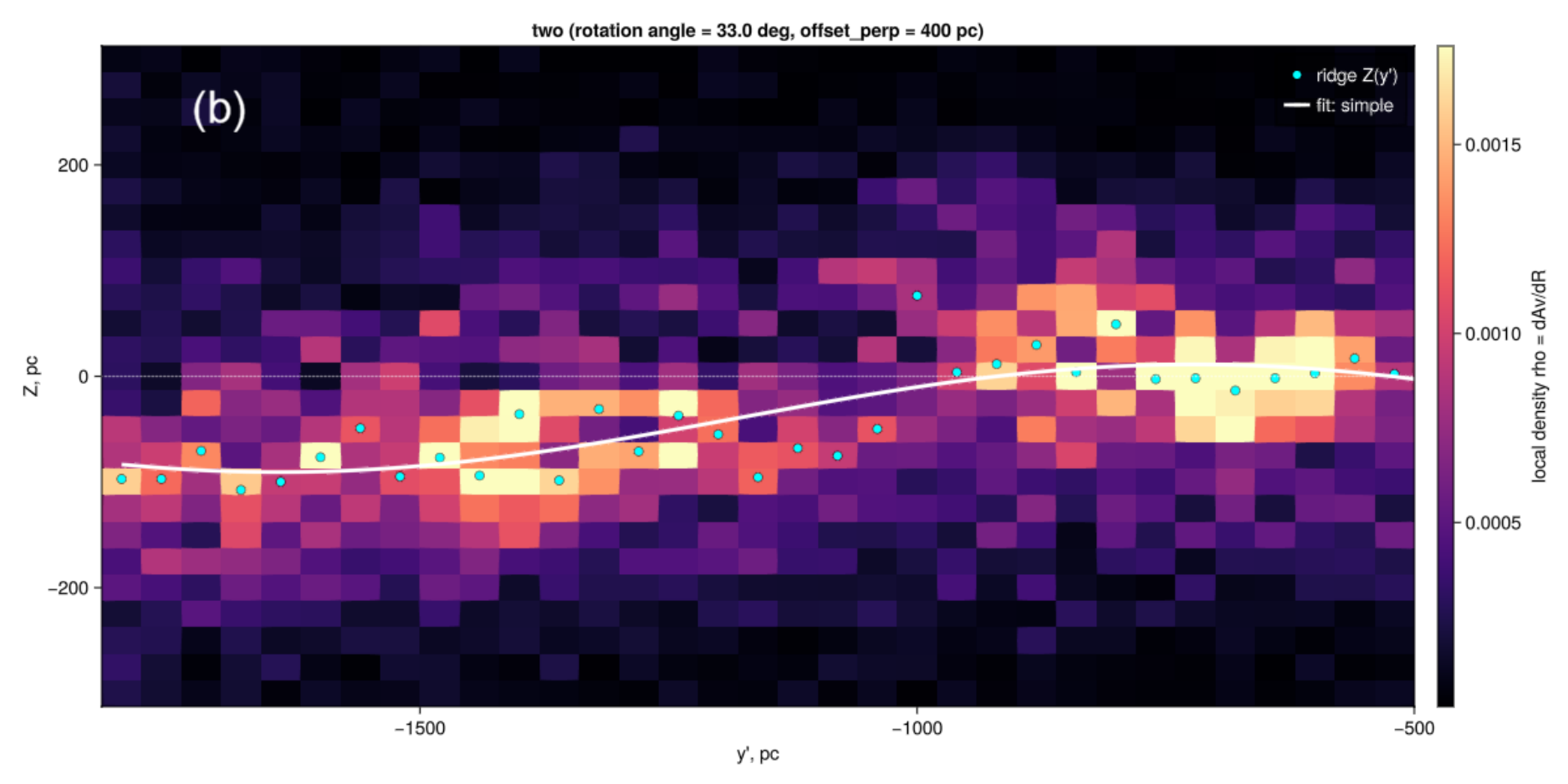}
\caption{Maps of the vertical distribution of dust in the Malpolon+Natrix~(a) and
Vela Ridge~(b) supercloud selection zones. }
 \label{f-xy-77}
\end{center}}
\end{figure}

 \section{RESULTS AND DISCUSSION}\label{rezult}
  \subsection{Radcliffe Wave, Split and Sagittarius Spur Extension}\label{three}
Fig. \ref{f-xy-3} shows the selection zones for the three most prominent structures on the map -- Radcliffe Wave, Split, and Sagittarius Spur Extension. The following three bands were considered here: Radcliffe Wave $(\alpha,x_{\perp,0},\Delta x_\perp,y'_{\min},y'_{\max},Z_{\max})=(31.4^\circ,-280,180,-2100,1800,350)$, Split $(40.9^\circ,135,220,0,1300,300)$, and Sagittarius Spur Extension $(35.85^\circ,980,260,-550,1050,300)$; All linear quantities here are given in pc, and the height range used was $|Z|<Z_{\max}$. Within each band, a map $\overline{\rho}_A(y',Z)$ was constructed with steps $\Delta y'=40$~pc and $\Delta Z=25$~pc, averaged over five cross-sections.

Fig. \ref{f-xy-4-RW} shows a smoothed map of the dust distribution in the vertical plane, constructed in the Radcliffe wave selection zone. The sinusoidal dependence found is shown, with a wavelength of about 2 kpc and a peak-to-peak amplitude of about 120 pc.

Figure \ref{f-xy-5} shows smoothed maps of the dust distribution in vertical planes, constructed in the selection zones of the Split (Fig. {f-xy-5}a) and Sagittarius Spur Extension ((Fig. \ref{f-xy-5}b) superclouds). Periodic dependencies are shown, which in both cases are of a small-scale nature, i.e., have wavelengths less than 0.5 kpc, and amplitudes less than 30 pc. All this is in good agreement with the result of the work of Kormann et al. (2026).

  \subsection{Anguis, Malpolon, Natrix and Vela Ridge}\label{four}
A fairly dense and extensive dust concentration is clearly visible in the Vela Ridge region on the f-xy-maps. Therefore, we created a selection zone for this supercloud, which, for negative values of $Y$, is very close to its outline in the paper by Kormann et al. (2026).

In the Anguis region, Kormann et al. (2026) did not observe high-amplitude (with amplitudes greater than 30 pc) long-wavelength (with wavelengths greater than 2 kpc) oscillations of vertical coordinates. The Anguis supercloud is not visible as a separate elongated structure on our Fig.~\ref{f-xy-map}. Therefore, we do not conduct a separate analysis of this structure in this paper.

But three extended, practically parallel to each other structures Anguis, Malpolon, Natrix (numbers 1, 2 and 3 on the map \ref{f-1}) on our maps~\ref{f-xy-map} are visible in the form of several clamps. However, the entire elongated structure is visible on the maps~\ref{f-xy-map}. Note that a similar single feature is also visible on the new dust distribution map in the work of Zhang et al.~(2026), for example, in Fig.~12 in the work of these authors, as well as on a similar map in the work of Lallement et al. (2022). Therefore, here, to study the nature of the vertical behavior of the coordinates, we formed a selection region shown in Fig. \ref{f-xy-66} and which we further denote as Malpolon+Natrix.

Fig. \ref{f-xy-66} shows a smoothed map of dust distribution in the $XY$ plane. One color scale on the right shows the depth of the map, and the second scale shows the depth of the vertical waves.

Fig. \ref{f-xy-77} shows smoothed maps of dust distribution in vertical planes, constructed in the selection zones of the Malpolon+Natrix~(a) and Vela Ridge~(b) superclouds. Periodic dependencies are shown,

Table ~\ref{tab:wavefit} shows the sinusoidal fit parameters found for four superclouds.
The table shows that three regions—Radcliffe Wave, Malpolon+Natrix, and Vela Ridge—have periodic perturbations in the vertical coordinates with wavelengths ranging from 2.5 kpc (Radcliffe Wave) to 2 kpc (Malpolon+Natrix and Vela Ridge).
We did not find similar long-wavelength and high-amplitude oscillations in the vertical coordinates in the Sagittarius Spur Extension and Split regions. The amplitude of the oscillations in the Split structure is less than 20 pc, so we do not present the characteristics of its waveform.

 \section{DISCUSSION}\label{discussion}
We note a difference between our processing method and the approach of Kormann et al. (2026). In their work, the input data was the 3D map of Edenhofer et al. (2024), where the logarithm of absorption was reconstructed using a Gaussian process with a priori smoothness conditions. Therefore, smoothing was essentially part of the 3D map construction procedure itself. Subsequent supercloud extraction was performed not by smoothing the 2D projection, but by the HOP algorithm in the 3D density distribution. This algorithm selects voxels above a given density threshold, identifies local maxima, and merges or separates adjacent structures based on the threshold, minimum cell count, peak density, and bridge density parameters. In contrast, in the present work, the 2D map $\Sigma_A(X,Y)$ was smoothed using an explicitly specified Gaussian kernel, and the Radcliffe wave and Split were then extracted as two predetermined extended regions. This approach is simpler and less automated, but it is useful for checking whether these structures are evident in the independent map of Gontcharov et al. (2025). The parameters for the Sagittarius Spur Extension should be considered as an additional control estimate, since a sinusoidal approximation for this structure was not provided by Kormann et al. (2026).

As follows from Fig. \ref{f-xy-5}, the Split region does not exhibit significant systematic perturbations of the vertical coordinates, while the nearby Radcliffe Wave region is subject to strong vertical perturbations (Fig.~\ref{f-xy-4-RW}). All this is in good agreement with the results of the analysis of Kormann et al. (2026) and is also of great significance in light of the rethinking of the evolutionary status of the Gould Belt (Gonz\'alez et al. 2026).

 The Gould Belt (Gould, 1874) appears on the celestial sphere as a wave-like chain of young objects -- OB stars, young open star clusters, a number of OB associations, and gas-dust clouds (Bobylev, 2014). Over the past few decades, the Gould Belt has been conceptualized as a spatial structure in the form of a disk, sometimes a donut, located at an angle of about 20 degrees to the galactic plane (e.g., Torra et al., 2000; Perrot and Grenier, 2003; Bobylev, 2006; Gontcharov, 2009). The disk radius is about 250 pc, and its center is located near the Sun (no further than 100-150 pc). A number of authors (e.g., Lindblad, 2000; Bobylev, 2006) note that this structure exhibits such kinematic effects as general expansion and proper rotation. According to this concept, the entire structure is assumed to have formed approximately 30–50 million years ago and to be evolving as a single entity. However, Gonz\'alez et al. (2026) reject the Gould Belt model as an evolving single entity. These authors argue that the spatial arrangement of OB associations in Scorpius-Centaurus and Orion forms an asterism. This accounts for the currently observed properties of the Gould Belt, but the system is not evolving as a single entity. The fact that the Orion OB association belongs to the Radcliffe Wave supercloud, and the Scorpius-Centaurus OB association belongs to the Split supercloud, is the main argument in support of this scenario.

 	\begin{table}[t]
		\caption{Parameters of sinusoidal approximation}
		\label{tab:wavefit}
		\begin{center}
			\begin{tabular}{lrrrrrr}
				\hline
	Structure &	$A$ & $\lambda$ &	$\varphi$ & $d$ &	$\varepsilon_a$  & $\varepsilon_\omega$ \\	
 	                   &	 pc   &          pc     &  rad &        pc &	$10^{-3}$ pc$^{-1}$ & 	$10^{-3}$ pc$^{-1}$ \\		\hline

  		Radcliffe Wave                    & 60 & 2581 &     0.620 &   $-53$ &    0.054  & $-0.262$ \\
		Malpolon+Natrix                  & 38 & 2029 & $-2.698$ &      83  &     0.365  &  0.510  \\
		Vela Ridge                           & 51 & 1841 & $-2.250$ &  $-40$ &        ---     & ---     \\
		Sagittarius Spur Extension & 24 &   412 &      0.053 &   $-7$  & $-0.650$ & $-0.115$ \\
				\hline
			\end{tabular}
		\end{center}
	\end{table}

The parameters of vertical perturbations of coordinates and velocities in the Radcliffe Wave region have been studied, for example, in the works of
Alves et al. (2020), Bobylev et al. (2022; 2026), and Konietzka et al. (2024). The estimates of the Radcliffe Wave parameters obtained in this study are generally in agreement with those of other authors. However, it should be noted that the wave amplitude estimates are highly dependent on the analytical methods used. After all, the wave here is not monochromatic; its amplitude varies depending on the $Y'$ coordinate, and in the solar region, the wave amplitude can reach values of 120--150 pc, as can be seen in Fig.~\ref{f-xy-4-RW}.

Maps of the distribution of dust matter between segments of the Perseus and Carina-Sagittarius spiral arms (e.g., Fig. 13 of Lallement et al. 2022) reveal the presence of low-amplitude ripples in the vertical direction, traced in various directions along the line of sight (from the position of the Sun). Of great interest, however, is the fact that, in addition to small-scale low-amplitude ripples, the Local System contains at least three significant structures with long-wavelength and high-amplitude oscillations of the vertical coordinates.

While the coordinate and kinematic properties of the Radcliffe Wave are fairly well established, the kinematics of the Malpolon+Natrix and Vela Ridge structures as individual entities has remained virtually unstudied. It would be interesting to confirm their presence, for example, using young open clusters (for which distance, age, and spatial velocity estimates are available) and to evaluate their kinematic properties.

  \section{CONCLUSION}\label{concl}
A detailed study of dust distribution in the Local System was conducted. Using the latest map by Goncharov et al., smoothed dust distributions were obtained, projected onto the galactic $XY$ plane, using various smoothing parameters. Within the 2-kpc radius region around the Sun, key structural features associated with the Radcliffe Wave, Split, Sagittarius Spur Extension, Malpolon+Natrix, and Vela Ridge superclouds are clearly visible.

In contrast to the work by Kormann et al. (2026), in the map by Gontcharov et al. (2025), the extended, nearly parallel structures of Anguis, Malpolon, and Natrix are visible only as a few clamps. However, one extended feature stands out in this region. To study the nature of the vertical behavior of the coordinates here, we formed a selection zone, which we designated Malpolon+Natrix. This approach is consistent with a similar elongated feature in the dust distribution map, for example, in the work by Zhang et al. (2026).

It is shown that in the Radcliffe Wave, Malpolon+Natrix, and Vela Ridge regions, there are periodic perturbations of the vertical coordinates with a wavelength from 2.5 kpc (Radcliffe Wave) to 2 kpc (Malpolon+Natrix and Vela Ridge) with oscillation amplitudes greater than 30 pc. In the Sagittarius Spur Extension and Split regions, similar long-wavelength and high-amplitude oscillations of the vertical coordinates were not detected.

We conclude that the membership of the Orion OB association in the Radcliffe Wave supercloud and the membership of the Scorpius-Centaurus OB association in the Split supercloud is an important argument in support of the proposal by Gonz\'alez et al. (2026) that the Gould Belt does not evolve as a single entity, but forms a so-called asterism.

It can be seen that in the Local System, in addition to small-scale, low-amplitude ripples, there are at least three significant, nearly parallel structures with long-wavelength, high-amplitude oscillations of vertical coordinates. In the future, it will be interesting to determine what might cause such disturbances.

\medskip

 \bigskip{BIBLIOGRAPHY}\medskip{\small
\begin{enumerate}

 \item
J. Alves, C. Zucker, A. A. Goodman, et al., Nature {\bf 578}, 237 (2020).

 \item
F. Anders, A. Khalatyan, A. B. A. Queiroz, et al., Astron. Astrophys. {\bf 658}, A91  (2022).

 \item
M. Barbillon, A. Recio-Blanco, P. de Laverny, and P. A. Palicio,  arXiv: 2511.12156 (2025).

 \item
V. V. Bobylev, Astron. Lett. {\bf 32}, 816 (2006).

 \item
V. V. Bobylev, Astrophysics {\bf 57}, 583 (2014).

 \item
V.V. Bobylev, A.T. Bajkova, Astron. Lett. {\bf 40}, 783 (2014).

\item
V.V. Bobylev, A.T. Bajkova, and Yu.N. Mishurov, Astron. Lett. {\bf 48}, 434 (2022).

 \item
V.V. Bobylev, N.R. Ikhsanov, A.T. Bajkova, Astron. Rep. {\bf 69}, 786 (2025).

 \item
V.V. Bobylev,  A.T. Bajkova, N.R. Ikhsanov, Astrophys. Bull. {\bf 81}, Issue~3 (2026).

\item
Gaia Collaboration,  T. Prusti, J. J. de Bruijne, A. G. A. Brown, et al., Astron. Astrophys. {\bf 595}, 1 (2016). 

\item
Gaia Collaboration, A.G.A. Brown, A. Vallenari, T. Prusti, et al.,
 Astron. Astrophys. {\bf 616}, 1 (2018). 

 \item
Gaia Collaboration, A. Vallenari, A.G.A. Brown, T. Prusti,
 et al., Astron. Astrophys. {\bf 674}, 1 (2023).  

 \item
G. A.  Gontcharov, Astron. Lett. {\bf 35}, 780 (2009).

 \item
G. A. Gontcharov, A. V. Mosenkov, MNRAS {\bf 472}, 3805 (2017).

 \item
G. A. Gontcharov, A. V. Mosenkov, MNRAS {\bf 475}, 1121 (2018).

 \item
G. A. Gontcharov, A. V. Mosenkov, MNRAS {\bf 483}, 299 (2019).

 \item
G. A. Gontcharov, A. V. Mosenkov, MNRAS {\bf 500}, 2590 (2021a).

 \item
G. A. Gontcharov, A. V. Mosenkov, MNRAS {\bf 500}, 2607 (2021b).

 \item
G. A. Gontcharov et al., Astron. Lett. {\bf 48}, 578 (2022).

 \item
G. A.  Gontcharov, A. A. Marchuk, S. S. Savchenko,  et al.,
Res. Astron. Astrophys. {\bf 25}, No 12, id. 125016  (2025).

   \item
B. A. Gould, Proc. Am. Ass. Adv. Sci. {\bf 115} (1874).

 \item
G. M. Green, E. Schlafly, C. Zucker, et al.,
ApJ {\bf 887}, 93 (2019).

\item
G. Edenhofer, C. Zucker, P. Frank, et al., Astron. Astrophys. {\bf 685}, A82  (2024).

\item
Yu. N. Efremov, {\it Star formation sites in galaxies} (Moscow: Nauka, 1989.

 \item
R. Konietzka, A.A. Goodman, C. Zucker, et al.,  Nature {\bf 628}, 62 (2024).

 \item
L. A. Kormann, J. Alves, M. P. Gonz\'alez, et al.,
Astron. Astrophys. {\bf 706}, A161 (2026).

 \item
R. Lallement, C. Babusiaux, J. L. Vergely, et al., Astron. Astrophys. {\bf 625}, A135  (2019).

 \item
R. Lallement, J. L. Vergely, C. Babusiaux, and N. L. J. Cox, Astron. Astrophys. {\bf 661}, A147  (2022).

 \item
P. O. Lindblad, Astron. Astrophys. {\bf 363}, 154 (2000).

\item
H. Mineur, MNRAS {\bf 90}, 789 (1930).

\item
Ogorodnikov K. F., {\it Dynamics of stellar systems} (Oxford: Pergamon, ed. Beer, A. 1965).

\item
C. A. Olano, Astron. Astrophys. {\bf 121}, 295 (2001).

\item
M. Pantaleoni Gonz\'alez, J. Alves, C. Swiggum, and I. Niederbrunner, arXiv: 2604.13225 (2026).

\item
C. A.Perrot, I. A. Grenier, Astron. Astrophys. {\bf 404}, 519 (2003).

\item
M. F. Skrutskie, R. M. Cutri, R. Stiening, et al., Astron. J. {\bf 131}, 1163 (2006). 

\item
 J. Torra, D.  Fern\'andez, and F. Figueras, Astron. Astrophys. {\bf 359}, 82 (2000).

\item
J. Vall\'ee, Astrophys. Space Sci. {\bf 363},  id. 243 (2018).

\item
A. V. Veselova, I. I. Nikiforov, Res. Astron. Astrophys. {\bf 20}, Issue 12, 209 (2020).

\item
Y. Xu, J. J. Li, M. J. Reid, K. M. Menten, et al., Astrophys. J. {\bf 769}, Issue 1, id. 15 (2013).

\item
Y. Xu, C. J. Hao, D. J. Liu, et al.,
  Astrophys. J. {\bf 947}, Issue 2, id. 54 (2023).

\item
L. Zhang, B. Chen, F. Qin, G. Li, H. Yuan, and Y. Ren,	 arXiv: 2607.06016 (2026).

 \end{enumerate} }
\end{document}